\documentclass[fleqn,usenatbib]{mnras}

\usepackage{newtxtext,newtxmath}

\usepackage[T1]{fontenc}
\usepackage{ae,aecompl}

\def\grs{$\gamma-$ray~}

\usepackage{graphicx}	
\usepackage{amsmath}	
\usepackage{amssymb}	
\usepackage[normalem]{ulem}
\usepackage{color,soul}

\def\gsim{\mathrel{\raise.5ex\hbox{$>$}\mkern-14mu
             \lower0.6ex\hbox{$\sim$}}}

\def\lsim{\mathrel{\raise.3ex\hbox{$<$}\mkern-14mu
             \lower0.6ex\hbox{$\sim$}}}

\title[Accretion Disk MHD Winds and Blazar Classification]{ Accretion Disk MHD Winds and Blazar Classification}

\author[S. Boula et al.]{
Stella Boula,$^{1}$\thanks{E-mail: stboula@phys.uoa.gr}
Demosthenes Kazanas,$^{2}$
and Apostolos Mastichiadis$^{1}$
\\
$^{1}$Department of Physics, National and Kapodistrian University of Athens, Panepistimiopolis, GR 15783 Zografos, Greece\\
$^{2}$NASA Goddard Space Flight Center, Greenbelt, MD, United States\\
}

\date{Accepted XXX. Received YYY; in original form ZZZ}

\pubyear{2018}

\begin{document}
\label{firstpage}
\pagerange{\pageref{firstpage}--\pageref{lastpage}}
\maketitle

\begin{abstract}
The Fermi Gamma-Ray Space Telescope observations of blazars show a strong correlation between the spectral index of their gamma-ray spectra and their synchrotron peak frequency $\nu_{\rm{pk}}^{\rm{syn}}$; additionally, the rate of Compton Dominance of these sources also
seems to be a function of  $\nu_{\rm{pk}}^{\rm{syn}}$. In this work, we adopt the assumption that the nonthermal emission of blazars is primarily due
to radiation by a population of Fermi-accelerated electrons in a relativistic outflow (jet) along the symmetry axis of the blazar's accretion disk. Furthermore, we assume that the Compton component is related to an external photon field
of photons, which are scattered from particles of the magnetohydrodynamic (MHD) wind emanating from the accretion disk. Our results reproduce well the aforementioned basic observational trends of blazar classification by varying just one parameter, namely the mass accretion rate onto the central black hole.

\end{abstract}


\section{Introduction}

Blazars (Flat Spectrum Radio Quasars and BL Lacs) are types of Active Galactic Nuclei (AGN) with
relativistic jets pointing toward us. The relativistic motion of these jets amplifies
their radiation emission, which in general dominates their disk emission, making these
objects extremely bright and therefore visible at large redshifts, $z$.
The blazar jet spectra are non-thermal \emph{par excellence}, implying emission by
inherently relativistic electron distributions. They consist of two broad ``humps",
one that spans the radio to optical-UV (and occasionally the X-ray) bands and another one
that extends from X-rays to multi-GeV and on occasion to TeV $\gamma$-rays. The first
is thought to be due to synchrotron emission by the jet's non-thermal electrons and peaks
in the IR--X-ray band, while the second by inverse Compton (IC) scattering of the
synchrotron or external (from an altogether different source) photons.
Quasi-simultaneous multiwavelength variability is also observed but not always, suggesting that the broader
emission of both ``humps" spans a region larger than that over which the electrons, radiating at a specific band, cool. The size of the emitting region is also not well established; the typical size estimates obtained from temporal variability arguments are affected by relativistic
plasma motion and radiative cooling. It varies among sources and even among different
observations of the same object and depends on the processes employed to model the
emission in specific bands. A typical emission region, at least in FSRQs where line
emission is detected, is thought to be the broad line region \citep{Sbar12}, of order of
$R_{\rm{BLR}} \sim 10^3 - 10^4 \, r_{s_8}$ (where $r_{s_8} \simeq 3 \times 10^{13} (M/10^8 M_{\rm{\odot}})$ cm is the
Schwarzschild radius of the black hole of mass $M$). On the other hand, the apparently
thin GeV emission \citep{Marscher10}, suggests a location (at least of this emission) at
a distance as large as $\sim 10^6 r_{\rm{s_8}} \sim 10$ pc, considerably larger than $R_{\rm{BLR}}$.

One would think that the broad frequency range of blazar emission, their multi-scale
variability and the strong dependence of the observed luminosity on the bulk Lorentz factor
of the relativistic jets and the observers' line of sight relative to its motion, would
preclude a systematic study of the blazar properties. Indeed, fits to their individual
multi-wavelength spectra in terms of Synchrotron - Compton models, require no less than
seven parameters \citep{Ghis11}. However, even with the limited $\gamma$-ray data of
\emph{CGRO}, certain regularities became apparent and are now known as the ``Blazar
Sequence" \citep{Foss98}: Blazars become redder with increasing bolometric luminosity
$L_{\rm bol}$, in that their synchrotron peak frequency $\nu_{\rm pk}^{\rm syn}$ decreases as
$L_{\rm bol}$ increases; at the same time their Compton dominance (CD) (i.e. the ratio of their IC to synchrotron luminosities) increases and so do their $\gamma$-ray spectral indices.

The blazar sequence, established with 132 objects out of which only 33 were detected in high
energy $\gamma$-rays, was supplanted with the launch of \emph{Fermi} and the discovery
of more than 1000 $\gamma$-ray blazars in the 2nd and 3rd Fermi Blazar Catalogs
(2LAC, 3LAC) \citep{Acker11,Acker15}, half of which with measured redshifts. These
compilations provided novel correlations that replaced those of the original blazar sequence.
Most importantly, these correlations were found to be independent of the source luminosity,
implying that the underlying physics are related to dimensionless parameters. These
correlations relate the slope $\alpha_{\rm{\gamma}}$ of the \grs spectrum to the synchrotron
peak frequency $\nu_{\rm pk}^{\rm syn}$ (Figs 17 and 10 of 2LAC and 3LAC respectively) and the Compton dominance to $\nu_{\rm pk}^{\rm syn}$ (Fig 27 of 3LAC; see also \cite{Finke13}).

Various aspects of the blazar sequence have been put forward by a number of
investigators: \citet{Ghis98} suggested it was based on the different radiative cooling
suffered by electrons in the different classes of sources, assuming the same heating
process in all. In this respect, the more luminous FSRQs rely on the (observed) BLR
line emitting clouds  to isotropize a non-negligible fraction of the accretion disk
luminosity, which is then (inverse) Comptonized by the jet's accelerated electrons and  boosted by its relativisitic motion to provide their higher
luminosities. BL Lacs with weaker line emission, are therefore less luminous and exhibit
higher individual electron energies. The dependence of the blazar sequence on $L_{\rm
bol}$, prompted \cite{Nieppola08} to suggest that it is the result of different
observer orientations and therefore different values of the beaming factor $\delta$. Pursuing a different point of view, \citet{Ghis09,Ghis11}
suggested that at small (normalized) accretion rates, transition of the accretion flow
to an ADAF \citep{NY95} state would cause removal of the BLR region, absence of the
concomitant cooling and reduction of the \grs slope to $\alpha_{\rm{\gamma}} <1$, thus
`dividing' blazars into FSRQs and BL Lacs on the basis of this parameter.
More recently, noting that the dependence of the blazar sequence on the
(relativistically boosted) blazar bolometric luminosity sounds nonphysical,
\citet{Ghis17} proposed an altogether different scheme, separating the BL Lac and FSRQ
behavior, with the former following the well defined blazar sequence, and the latter
forming their own distinct class.

\indent The present work is motivated by observations and models of an altogether different aspect of AGN phenomenology, namely by features that have been termed ``Warm Absorbers" (WA). 
These are blue-shifted absorption features in the X-ray spectra, indicating the presence of ionized outflows, a ubiquitous feature both in the UV and X-ray AGN spectra \cite{Cren2003}.

The presence of absorption features is very important for the determination of the properties of these outflows since each such transition ``lives" at a well defined value of the photo-ionization parameter $\xi \; (= L/n(r)r^2$, where $L$ is the ionizing luminosity, $n(r)$  the plasma density and $r$ its distance from the black hole).  
The entire wind density profile of a transition can be determined by measuring both $\xi$ and the absorption depth of said transition, along the line of sight.  This was the approach of \citet{Behar09} who, assuming a density profile of the form $n(r) \propto r^{-p}$, fitted the multiple absorber data to obtain values of $p$ close to 1 ($1 < p \lsim 1.22 $). It is important to note that these density profiles are much
shallower than those expected in radiation driven winds. Such profiles are naturally
produced in self-similar accretion disk MHD winds \citep{CL94,FKCB}, launched over the
entire disk domain, spanning a radial range of $R_D \sim 10^5-10^6 \; r_{\rm{s_8}}$. At the same time the winds have a very steep density dependence on the polar angle
$\theta$. 
The steep $\theta$ dependence gives to these winds a toroidal appearance, considered for this reason to be the ``dusty tori" invoked in AGN unification \citep{AntonMill,KK94}. The shallow radial density profile of these winds is of importance in the determination of the blazar
radiative properties: it allows the scattering of the accretion disk's continuum radiation
over large distances, so that it becomes roughly isotropic over radii comparable to the
extent of the disk ($r \sim R_D$). This radiation will then appear in the forward direction on the frame of the relativistic jet out to distances $r_z$ along its axis, such that $r_z/\Gamma  \sim  R_D $ ($\Gamma$ is the bulk Lorentz factor). It can then be scattered efficiently by the high energy electrons of the
relativistically moving jet to produce its observed X-ray and \grs radiation; these are
the photons employed in the External Compton (EC) scattering that appear to produce the dominant component of the \emph{Fermi} blazar spectra. 
The density of these photons is related to the mass accretion rate $\dot{m}$, normalized to the Eddington rate. Assuming the magnetic field to be in equipartition with the kinetic energy density,
it can be shown that the magnetic energy density is a factor of $\dot m$ (or $\dot m^2$ in the ADAF case) larger than that of the continuum luminosity isotropized through scattering in the wind.   
However, the latter will be larger than the former by a factor
of $\Gamma^2$ on the jet frame. Therefore, it is expected that the IC will be larger
than the synchrotron component when $\dot m \Gamma^2 >1$, providing a vestige of the blazar LAT correlations in terms of a single parameter, namely $\dot m$. \\
\indent The present work provides a detailed study of this argument. In \S \ref{s2} we outline the model in more detail, with particular emphasis on the wind structure, the radiation and magnetic fields and the shape of electron distribution function. In \S \ref{s3} we compare the results of our
models with the LAT blazar correlations, and we summarize and discuss our results in \S \ref{s4}. 
\section{Modelling the Blazar Sequence}\label{s2}
\indent Previous studies \citep{Foss98} have shown the existence of a Blazar classification, known as Blazar Sequence, which presents a connection between the observed bolometric luminosity and the Spectral Energy Distribution (SED) of blazars. In general, the blazar SED is modeled with a simple one-zone leptonic model  \citep{Finke13}. In this case, accelerated electrons produce a low energy component through synchrotron radiation and a high energy one through inverse Compton scattering. For the latter component, the origin of the target photon field can be either internal (from synchrotron radiation) and/or external (\cite{Sikora94}, \cite{DSM}). 

In the present paper we will also  employ this one-zone leptonic model; however, before doing so,  we will present simple scalings of all the important parameters on just two fundamental physical quantities of the problem, namely the mass of the black hole and the mass accretion rate.

\subsection{Scaling of the physical parameters }
\indent We begin by calculating the specifics of the external photon field. This is based on the self-similar MHD accretion disk wind model of \cite{FKCB} and the wind is described by steady-state, non-relativistic equations, which include gravity and pressure. These authors derive a density profile for the wind particles of the form 
\begin{equation}\label{eq:nr}
 n(r,\theta) = n_0 (r_s/r)^p \, e^{5(\theta -\pi/2)},
\end{equation}
 where $n_0=\frac{\eta_w  \dot{m}}{2\sigma_{\rm T}r_{\rm s}}$, $r_{\rm s}$ is the
Schwarzschild radius, $\sigma_{\rm T}$ is the Thomson cross section, $\dot{m}$ is the normalized mass accretion rate ratio to the Eddington one and $\eta_{\rm w}$ is the ratio of the mass outflow rate of the wind to the mass accretion rate $\dot{m}$, which is assumed to be $\eta_{\rm w} \simeq 1$. The value of the index $p$ is between 1 and 1.2 \citep{FKCB17,FKCB18}; here for simplicity we will assume $p=1$, while different values will be explored in a future publication.\\
\indent As a first order approximation we can ignore the dependence of $n(r,\theta)$ on the polar angle $\theta$. In that way one is able to calculate analytically the optical depth 
of the wind particles to
disk photons scattering. Setting therefore at Eq. (\ref{eq:nr}) $p=1$ and assuming a region between any radii $R_1 $ and $R_2$ (as they are measured from the central object), the Thomson optical depth can be written as 
\begin{equation}
\tau_{\tau}(R_1,R_2)=\int_{R_1}^{R_2} n(r)\sigma_{\rm T}{\rm{d}r }= n_0 \sigma_{\rm T}r_s \ln ({R_2}/{R_1}).
\end{equation}
\indent In order to estimate the energy density of the scattered photon field we should define first the luminosity of the accretion disk:
\begin{equation}
    L_{\rm disk}=\left\{
                \begin{array}{ll}
                  \epsilon \dot{m}{\cal M}L_{\rm Edd}  ~~~&{\rm for} ~\dot{m}\gsim 0.1,\\
                  \epsilon \dot{m}^2{\cal M}L_{\rm Edd}~~&{\rm for} ~\dot{m} \lsim 0.1,  \\
                \end{array} \right.
\end{equation}
where $L_{\rm Edd}=1.28 \times 10^{38}~\rm{erg/sec}$ is the Eddington luminosity of one solar mass and ${\cal M}= M_{\rm BH}/M_{\rm \odot}$, where $M_{\rm BH}$ is the mass of the black hole and $\epsilon$ is the efficiency of the conversion of the accreting mass into radiation. The case of $\dot{m}\geq 0.1$ corresponds to a radiatively efficient disk and the case of $\dot{m}< 0.1 $ to a disk in an ADAF state.  
\\
\indent Therefore in the case where $\tau_{\rm T} \ll 1$ and $R_1 \ll R_2$ the photon energy density of the scattered photons is 
\begin{equation}\label{eq:sc}
  U_{\rm sc}= \frac{L_{\rm disk}\tau_{\rm T}}{4\pi R_2^2 c} ~.
\end{equation}
We will use this value as the external photon energy density when applying the one-zone leptonic model.
\\
\indent The other important quantity, i.e. the magnetic field strength, can be defined from the accretion power of the source, assuming that it is in some state of equipartition with the kinetic energy density of the accreted mass. Thus, if we define the accretion power of the source as:
\begin{equation}
P_{\rm acc}= \dot{m}{\cal M}L_{\rm Edd},
\end{equation}
then the energy density of the magnetic field at the base of the outflow can be written as
 \begin{equation}
  U_{\rm B_0}=\frac{\eta_b P_{\rm acc}}{4\pi (3r_{\rm s})^2c},
 \end{equation}
where $\eta_{\rm b}$ is a proportionality constant ($\eta_{b}<1$) and assuming an inner disk radius of $3r_{\rm s}$. 
Furthermore, in order to determine the strength of the magnetic field along the axis of the jet we assume that
\begin{equation}\label{eq:B}
  B=B_0 ({z_0}/{z}),
\end{equation}
where $B_0$ is the value of the magnetic field at the base of the jet, $z$ the distance from the central engine and $z_0=3 r_{\rm s}$ . 

\subsection{The radiation model}
\label{sec:rad_model}

 \begin{figure}
   
  \centering
     \includegraphics[width=0.42\textwidth,trim={0 0.5cm 0 0}]{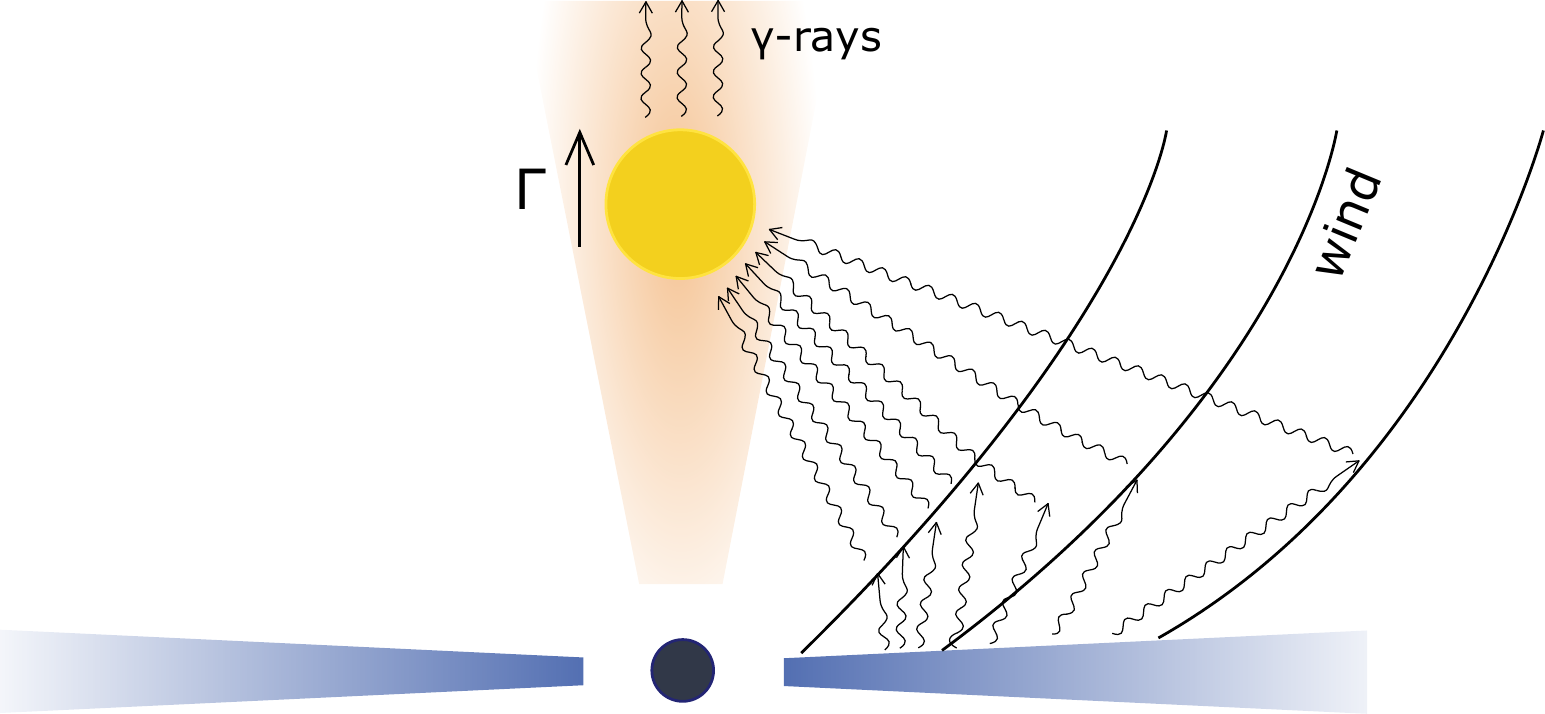}
  \caption{Schematic of the disk wind model for the $\gamma$-ray production (not in scale). Disk photons are scattered on wind particles and create an isotropic external photon field for the relativistic blob shown 
  in yellow, $\Gamma$ is the bulk Lorentz factor.}\label{fig:model}
\end{figure}
Having defined the external photon density and magnetic field, one can calculate the SED of the source in the case of a simple one-zone leptonic model. We assume, therefore, that the emission of the relativistic jet is dominated by a stationary single zone at some height z along the jet axis. The emitting region is assumed to be spherical with a radius $R_{\rm b}$ in its co-moving frame and it moves with a highly relativistic speed $\beta c$, giving it a Lorentz factor $\Gamma= (1-\beta^2)^{-1/2}$. Its velocity makes an angle $\phi$ to the line of sight, so the Doppler factor is $\delta =[\Gamma(1-\beta\cos \phi)]^{-1}$.\\
\indent Relativistic electrons are injected into the spherical volume and lose energy through synchrotron radiation on the embedded magnetic field of strength $B(z)$ (given by Eq. (\ref{eq:B}))  and inverse Compton scattering on internal and external photons.

Figure (\ref{fig:model}) gives a schematical representation of the isotropization of the disk photons on the wind and their consequent upscattering to $\gamma$-rays by the relativistic electrons.
\\  
\indent Furthermore, we assume that electrons are injected in the source with a broken power law distribution given by 
 \begin{equation}\label{eq:e}
    Q_{\rm e}=\left\{
                \begin{array}{ll}
                 k_{\rm e_1} \gamma^{-s} ~~~~~~~~~~~~~\rm{for} ~\gamma_{\rm min}\leq\gamma\leq \gamma_{\rm br},\\
             k_{\rm e_2} \gamma^{-q}e^{-\gamma/\gamma_{\rm max}}~~\rm{for} ~\gamma_{\rm br}\leq \gamma \leq\gamma_{\rm max},
                \end{array}
              \right.
 \end{equation} 
where $L_{\rm inj}^e=m_{\rm e}c^2\int_{\gamma_{\rm min}}^{\gamma_{\rm max}} Q_{\rm e}(\gamma) \gamma {\rm{d}}\gamma=\eta_{\rm e} P_{\rm acc}$, where $\eta_{\rm e}$ is a proportionality constant. Here, $\gamma_{\rm min}$ and $\gamma_{\rm max}$ are the minimum and maximum electrons Lorentz factors,  $s $ and $q$ the indices for the electrons' distribution before and after the break respectively,
$ \gamma_{\rm br}= \frac{3m_{\rm e}c^2}{4\sigma_{\rm \tau}ct_{\rm dyn}U_{\rm tot}}$, where $t_{\rm dyn}=\frac{R_{\rm b}}{c}$ is the dynamic time scale, while $U_{\rm tot}$ is the total energy density that is given by $U_{\rm tot}=U_{\rm B}+U_{\rm ext}+U_{\rm SSC}$,
where $U_{\rm B}$ is the energy density of the magnetic field at position z, $U_{\rm ext}=\Gamma^2 U_{\rm sc}$ is
 the energy density of the scattered photons as measured in the co-moving frame and $U_{\rm SSC}$ is the energy density of the produced synchrotron photons, which is calculated self-consistently from the numerical code. In order to calculate the SED of scattered photons we have assumed that the disk emits like a black body with a characteristic temperature $T_{\rm disk} $.
 We also assume that $\gamma_{\rm max} \gg \gamma_{\rm br}$ and for concreteness we use
$\gamma_{\rm max}=10^{4}\gamma_{\rm br}$.\\
 \section{Results}\label{s3}
 
\begin{figure}
  \centering
     \includegraphics[width=0.54\textwidth,trim={0 0.7cm 0 0.7cm}]{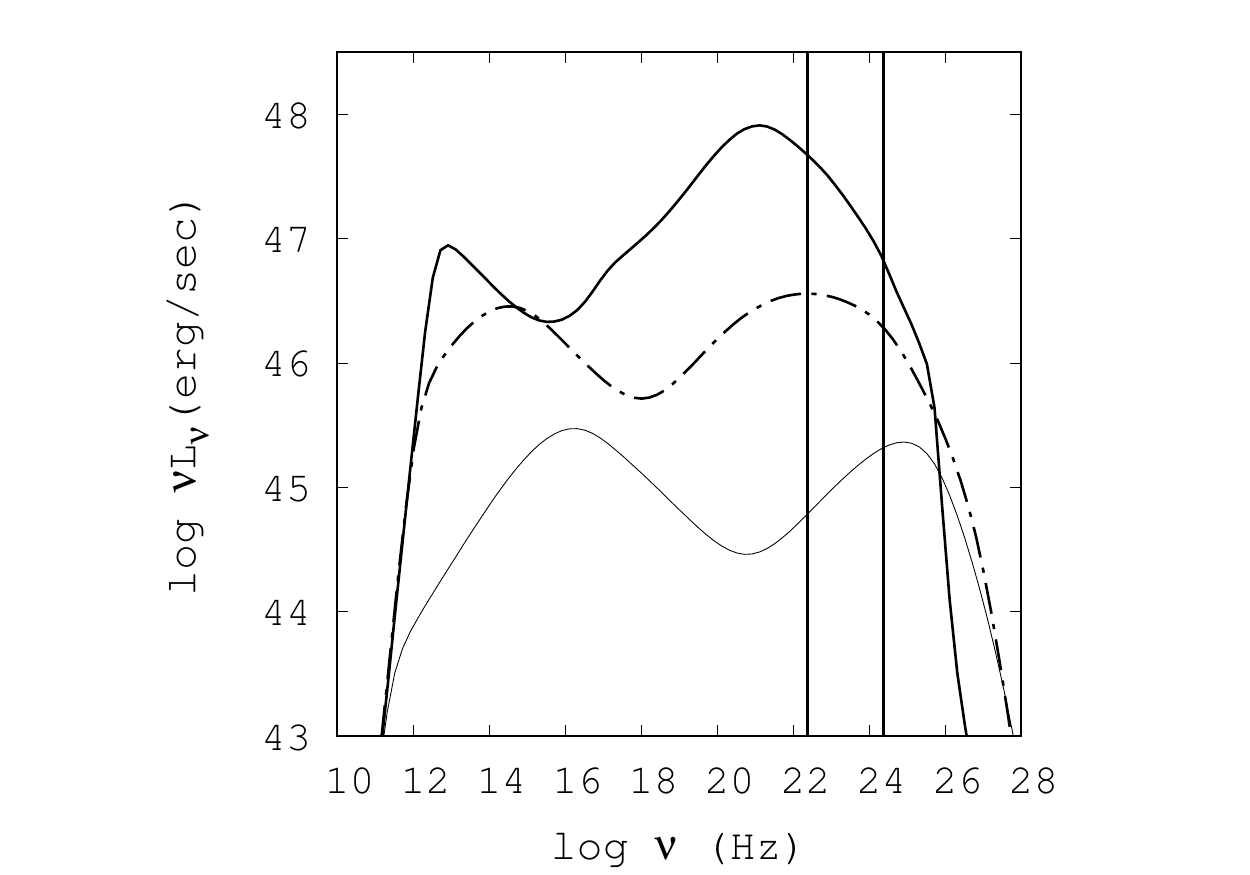}
  \caption{Multiwavelength spectra are calculated for different mass accretion rates. The straight, dotted, straight-dotted line  refers to $\log\dot{m}=-0.5$, $\log\dot{m}=-1.5$ and $\log\dot{m}=-2.5$ respectively. The black hole mass is $\mathcal{M}=10^9$ and the parameters' values are given in Table \ref{tab1}. The vertical lines show the Fermi energy band. }\label{a}
\end{figure}

\indent In order to calculate the blazar SED we solve the kinetic equations of electrons and photons as described by \cite{MK95}. As we pointed at the previous section, all  input parameters  required for the calculation of the spectrum are scaled with $\dot{m}$ and ${\cal{M}}$.
Thus using Eqns (\ref{eq:sc}), (\ref{eq:B}) \& (\ref{eq:e}) one can write $
 U_{\rm B} \propto \frac{\dot{m}}{\cal{M}}$,
  $U_{\rm ext}\propto U_{\rm sc}\propto \frac{\dot{m}^{\alpha+1}}{\cal{M}}~~(\alpha =1 ~{\rm{for}}~\dot{m}\geq 0.1 ~~{\rm{and}}~ \alpha=2 ~{\rm{for}}~\dot{m}<0.1)$,
 $\gamma_{\rm br}\propto \dot{m}^{-1} (1+\dot{m}^{\alpha})^{-1}$ and 
 $ L_{\rm e}^{\rm inj}\propto\dot{m}\cal{M}$. \\
\indent Furthermore, following \cite{Ghis09} we relate the mass accretion rate to the blazar type. Thus, we assume that the case $\dot{m} > 0.1$ corresponds to a radiatively efficient disk and therefore to an FSRQ object. On the other hand, the case $\dot{m} \ll 0.1$ corresponds to a disk in an ADAF state and thus corresponds to a BL Lac object.
Furthermore, BL Lac objects are divided in three subcategories, LBL, IBL and HBL, which refer to a low, intermediate and a high synchrotron peak frequency respectively. We model these by using decreasing values of $\dot{m}$. \\
\indent Table \ref{tab1} shows the values of the calculated input parameters by varying only the mass accretion rate $\dot{m}$. 
In this example we have kept the mass of the black hole constant. Figure \ref{a} depicts the obtained multiwavelength spectra and, as it can be seen, our simple, staightforward scalings reproduce the trend of the observed Blazar Sequence. Here we have assumed constant values for $\Gamma$ and $\delta$ factors. For these cases we use a constant temperature for the accretion disk. In the case of high  $\dot{m}$, $U_{\rm ext}$ is also high and the source is Compton dominated and reproducing FSRQs objects. On the other hand, low values of $\dot{m}$ correspond to low $U_{\rm ext}$ and as a result SSC scattering starts dominating over EC scattering. In this case the SED of BL Lac objects is reproduced. \\
\indent Furthermore, the observed peak of the synchrotron component for electrons with Lorentz factor $\gamma_{\rm br} $ will appear at $\nu_{\rm pk}^{\rm syn}\propto \delta B \gamma^2_{\rm br}$. Using the relations described above, we find 
\begin{equation}\label{eq.n}
 \nu_{\rm pk}^{\rm syn} \propto {\cal{M}}^{-1/2}  {\dot{m}^{-3/2}}/({1+\dot{m}^{\alpha})^2}.
 \end{equation}
Therefore, while a low value of $\dot{m}$  corresponds to lower values of $B$, the corresponding increase in $\gamma_{\rm br}$ leads to a higher value for  $\nu_{\rm pk}^{\rm syn}$. On the other hand, as $\dot{m}$ increases, despite the $B$ increase, the decrease of $\gamma_{\rm br}$ shifts $\nu_{\rm pk}^{\rm syn}$ to lower frequencies. 

For a fixed observation band between $\nu \sim 10^{22} -10^{24}$ Hz, as that of the \emph{Fermi}-LAT, it is clear from Fig.\ref{a} that the above scalings produce a progressive hardening of the $\gamma$-ray spectra by the progressive decrease of their index, 
$\alpha_{\gamma}$, with the decrease of the (normalized) accretion rate $\dot m$.

\begin{figure}
 \centering
     \includegraphics[width=0.5\textwidth,trim={0 0.7cm 0 0.7cm}]{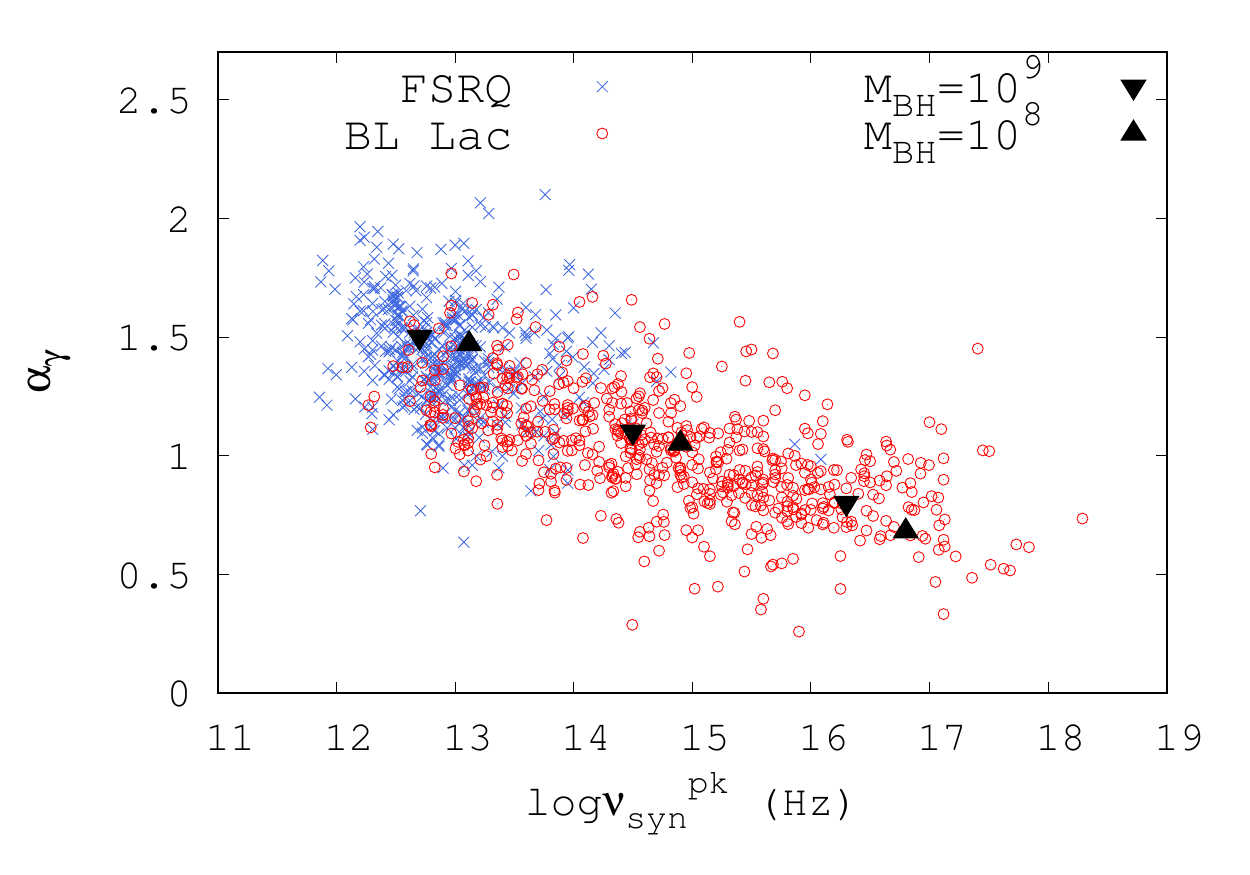}
  \caption{The variation of the \grs photon index $\alpha_{\gamma}$ as a function of the frequency of the synchrotron peak $\nu_{\rm syn}^{\rm pk}$ as shown in the 3LAC compilation.
The triangular points show our model results for the two values of the $\rm M_{\rm BH}$.
Three sets of points are produced for $\log \dot{m}$ values of -0.5, -1.5 and -2.5 (left to right).
}
\label{b}
\end{figure}

In Fig. \ref{b} we superimpose the results of our model calculations on the compilation of $\alpha_{\gamma} ~vs.~ \nu_{\rm pk}^{\rm syn}$ of 3LAC. 
We have chosen two values for the black hole mass, $\mathcal{M} = 10^8 ~\& ~ \mathcal{M} = 10^9$ and three values of $\log \dot m$, namely -0.5, -1.5 and -2.5. We have left all the other values associated with the emission, i.e. $\Gamma, \delta, s, q$, constant. We see that, as argued qualitatively, the parameter that sets the range of variation of these two observables is the dimensionless accretion rate $\dot m$. The proximity of points differing in mass by a factor of ten suggests that the mass is not the major driver of this relation (and the blazar main sequence generally.

\begin{table} 
\begin{center}
\resizebox{\columnwidth}{!}{
  \begin{tabular}{ c|c | c | c | c | c | c }
    \hline
    $ \dot{m}$ & $ B({\rm G}) $ & $U_{\rm ext}$ $\left(\frac{\rm erg}{\rm{cm}^3}\right)$ & $L_{\rm e}^{\rm inj}$ $\left(\frac{\rm erg}{\rm{sec}}\right)$ & $ \gamma_{\rm br}$& Blazar Class\\ \hline
         -0.5 & -0.3  & -1.4 &45.2 & 2.3 & FSRQ\\ \hline
     -1.5 & -0.8 & -4.6 & 44.2& 3.3  & LBL \\ \hline
    -2.5 & -1.3  & -7.6 &43.2& 6.5&  HBL \\    
    \hline
  \end{tabular}} 
\end{center}
\caption{Parameter values for different mass accretion rates in the case of $\mathcal{M}=10^9$. 
All the values are in a logarithmic scale.
The location of the source is at $z=1pc$, we assume a spherical
region for the external photon field between radii $R_1=9\cdot 10^{16}~\rm{cm}$ and $R_2=3\cdot 10^{19}~\rm{cm}$. The efficiencies of the mass accretion rate to the magnetic field, electrons luminosity and external photon field are $\eta_{\rm b}=0.1$, $\eta_{\rm e}=0.05$ and $\epsilon=0.5$ respectively. The indices for the electron distribution are $s=2$ and $q=2.75$. The bulk Lorentz factor is $\Gamma=30$ and the Doppler factor is $\delta=15$. The characteristic temperature of the disk is $T_{\rm disk}=3 \cdot 10^3 $ K. }\label{tab1}
\end{table}

\section{Summary-Discussion}\label{s4}

\indent In this Letter we have shown that a one-zone leptonic model with a simple scaling of its input parameters on the (dimensionless) mass accretion rate, $\dot{m}$, can reproduce the basic trends of the Blazar Sequence, as first suggested by \cite{Foss98} and further elaborated by the \emph{Fermi} 2LAC and 3LAC results. A key feature of our model is that the target photons necessary for the external Compton scattering are provided by the reprocessed disk radiation on a MHD wind. This is possible because the wind has a shallow density profile that results in a Thomson optical depth, which varies logarithmically with the outer disk radius but with normalization related directly to $\dot{m}$; additionally, the strength of the magnetic field is also related to $\dot{m}$, thus reducing the freedom of the relevant parameters. Furthermore, $\dot{m}$ determines also the efficiency of the disk luminosity, with a transition to an ADAF state for small values of $\dot m$. As a result, intrinsically powerful blazars, i.e. FSRQs, have larger ratios of external photon to magnetic field energy densities, and therefore are Compton dominated, while less intrisically powerful ones, like BL Lacs that are assumed to be in an ADAF state, are synchrotron dominated. 
Figure \ref{a} shows how the multiwavelength spectra of these sources evolve with varying $\dot{m}$ so that the corresponding \grs slope and $\nu_{\rm pk}^{\rm syn}$ vary to cover a region consistent with that of the \emph{Fermi} data (see Fig. \ref{b}). 

The basic aim of this work was to reproduce the theoretical Blazar Sequence using a minimum number of free parameters. The fact that the strongest blazar correlation produced by the Fermi data compilation (see Fig. \ref{b}) involves only spectral attributes, guided us to determine the most significant parameter. Note that, in order to calculate the SED of the sources it is necessary to specify some physical quantities such as the magnetic field strength $B$, the characteristics of the electrons distribution ($L_{\rm e}^{\rm inj},~\gamma_{\rm min}, ~\gamma_{\rm br},~\gamma_{\rm max} ~s,~q$) and the energy density of the external photon field $U_{\rm ext}$. 
These physical quantities turn out to be functions of $\dot{m}$ and the mass of the black hole, $\mathcal{M}$, (see \S \ref{s2}). From our numerical calculations we found that $\dot{m}$ is the parameter that has the greatest contribution in the reproduction of Fig. \ref{b}, while the other parameter, $\mathcal{M}$, does not play a key role. We mention, however, that low values of $\mathcal{M}$ can reproduce the multiwavelength spectra of FSRQs with low bolometric luminosity, \citep{Ghis17}. About the efficiencies $\eta_{\rm b}, ~\eta_{\rm e},~ \epsilon$ that also affect the physical quantities $B,~L_{\rm e}^{\rm inj}, ~ U_{\rm ext}$ respectively, preliminary calculations have shown that their variations can reproduce the spread of the observables of Fig. \ref{b} under the basic assumption of varying $\dot{m}$. Furthermore, we used ad-hoc values of the slopes before and after $\gamma_{\rm br}$ so as to be in agreement with the $\gamma$-ray photon indices in Fig. \ref{b}. Finally, we found that varying the Doppler factor $\delta$ has a similar effect as varying $\mathcal{M}$ and for this reason we kept its value constant.

Despite the fact that the results of this work are similar to \cite{Finke13}, the basic assumptions are different. In that paper the strength of the magnetic field is directly related to the energy density of the external photon field, while, in ours, quantities are related implicitly to $\dot{m}$.

To sum up, we have succeeded in  reproducing the theoretical Blazar Sequence by varying only one free parameter ($\dot{m}$) while keeping everything else fixed. We will examine systematically the effects of the secondary physical quantities in a future publication. 
\section*{Acknowledgements}
DK acknowledges support by NASA ADAP and  Fermi GI grants. We thank Dr. Maria Petropoulou for useful discussions and comments on the final manuscript. Also, we thank Dr. Stavros Dimitrakoudis for Fig. \ref{fig:model}.




\bibliographystyle{mnras}
\bibliography{general} 



\bsp	
\label{lastpage}
\end{document}